# Optical Multilayer Thin Film Structure Inverse Design: From Optimization to Deep Learning

Taigao Ma[1, +], Mingqian Ma[2], L. Jay Guo[2] *

[1]*Department of Physics, University of Michigan, Ann Arbor, MI 48109, USA*

[2]*Department of Electrical Engineering and Computer Science, University of Michigan, Ann Arbor, MI 48109, USA*

[+] *Work done while at University of Michigan. Currently at Visa Research, Austin, TX 78759, USA.*

* *Correspondence: guo@umich.edu*

## Summary

Optical multilayer thin film structures have been widely used in numerous photonic domains and applications. The key component to enable these applications is the inverse design. Different from other photonic structures such as metasurface or waveguide, multilayer thin film is a one-dimensional structure, which deserves its own treatment for the design process. Optimization has always been the standard design algorithm for decades. Recent years have witnessed a rapid development of integrating different deep learning algorithms to tackle the inverse design problems. A natural question to ask is: how do these algorithms differ from each other? Why do we need to develop so many algorithms and what type of challenges do they solve? What is the state-of-the-art algorithm in this domain? Here, we review recent progress and provide a guide-tour through this research area, starting from traditional optimization to recent deep learning approaches. Challenges and future perspective are also discussed.

## Introduction

Optical multilayer thin film structure[1–3] (shortened as multilayer structure) is a type of photonic platform that consists of multiple layers of material stacking together, with thickness typically in the range of tens of nanometers to several hundred nanometers. Because of their ease of fabrication, they have been widely used in many different photonic applications. For example, most photovoltaics[4,5] devices are based on thin film structures to achieve high power-conversion efficiency in large-scale manufacturing. Multilayer thin film-based transmissive/reflective filters[6,7] are widely used as functional coatings and protective layers for optical lenses and glass pieces. The light interference phenomenon inside multilayer structures also provides an ideal platform to make structural color coatings[8,9] as an eco-friendly and advanced decorations with high aesthetic expression. Other important applications include displays[10,11], touch screens[12], radiative cooling devices[13], to name a few.

For a given multilayer structure, obtaining its optical responses is easy and straightforward, through well-developed electromagnetic simulation algorithms, such as Transfer Matrix Methods[14] (TMM). However, the inverse problem, to design a multilayer structure with desired optical responses is non-trivial and usually does not have a closed-form. The inverse design requires determining the total number of layers and identifying the best combination of materials at each layer as well as their corresponding thickness, which is a combinatorial optimization problem. At the early stage, many designs are based on researchers' physics intuition or optical knowledge. However, this method is only suitable for some simple and conventional structures. As design complexity increases, optimization-based methods have been widely adapted to automatically identify optimal structures, including needle optimization[15,16], particle swarm optimization[17], genetic algorithms[18], etc. To speed up the design process, in recent several years, researchers have started to use deep learning as alternative approaches for inverse design[19–23]. After learning from the training dataset, these deep learning models can establish a direct mapping from optical responses to the multilayer structure, therefore, can finish the design process within a fraction of a second, with speed much faster than the optimization approach. Many different deep learning models have already emerged and demonstrated excellent performance in diverse applications, including tandem networks[19,24,25], Variational Auto-Encoders (VAE)[26], Generative Adversarial Networks (GAN)[27], Mixture Density Networks (MDN)[20,21], etc. There are also many nice benchmarking works that compare different inverse design methods[28–30].

With such significant progress in multilayer structure inverse design, a comprehensive review on these inverse design algorithms could benefit the relevant research community by providing a high-level perspective on the development and future directions. Though there are many excellent reviews that summarize deep learning-based inverse design for photonic and nano-photonic[31–38], there is essentially no review that focuses on the multilayer structure inverse design. Different from other photonic structures such as metasurface or waveguide, multilayer thin film is a one-dimensional structure, which requires special treatment and considerations during design process. For example, recently many researchers started to treat multilayer structure as a sequence[23,39,40], which is a special type of data that has not been considered in other types of photonic structures.

In this review, we will fill this gap by providing a perspective of multilayer structure inverse design and looking through the research roadmap. In the first part, we will provide an overview of multilayer structure inverse design by dissecting the representations of both multilayer structures and optical responses, and present the overall challenges. Then we will go through different mainstream inverse design methods and summarize their achievements in a historic perspective: starting from the traditional optimization-based methods over decades ago, then the deep learning using vectorized representation starting around several years ago, the recent combined methods using both optimization and deep learning, and finally the deep learning using sequential representation, which was just proposed in recent two years. We hope this review could provide insights on current research development and promote the exploration of new research frontiers.

## Multilayer Structure Inverse Design

In this section, we will provide a general introduction to multilayer inverse design fundamentals. Inverse design usually deals with finding the suitable structures that can satisfy desired optical response. Therefore, there are two aspects that researchers are interested in: multilayer structures and optical responses. We will first describe core concepts that we need to figure out when designing a multilayer structure, as well as three different methods when representing a multilayer structure. This is the part that readers can tell the difference between multilayer design and other nanophotonic inverse design. Following this, several common types of optical responses with general interest and wide applications will be discussed. In the last part of this section, we will summarize several challenges during inverse design and provide an illustrative table to show how existing inverse design methods can tackle these challenges in different aspects. Details of these methods will be discussed in the following sections.

### Design Parameters for Multilayer Structure

As the name suggests, multilayer structure consists of multiple layers of materials stacking on top of each other. Usually, each layer can be made from a different material with different thickness, and these different layers contribute together to control light propagation and interaction inside medium, and collectively determine the multilayer structure’s optical responses. For a multilayer structure, there are three sets of parameters to be determined during design process: the total number of layers, material combinations at each layer, and their corresponding thickness.

The total number of layers usually depends on the type of optical responses. For example, only 3-4 layers of suitable materials can already make a nice structural color filter[8,41,42]. However, in some other cases, such as designing a perfect reflective or transmissive filter, the total number of layers can easily exceed 100. Generally, there is no specific criterion for how many layers should be used for a given application; however, it is always desirable to use structures with fewer layers considering the fabrication complexity and cost. Material combinations is another important design parameter. In optical frequency, different materials usually have different refractive indices, which uniquely determines how light will pass through and interact with the layers. Based on their absorption ability in optical frequency, most materials can be divided into two types: dielectric, or metallic. Dielectric materials such as silicon dioxide ($SiO_2$) or titanium dioxide ($TiO_2$) typically have low optical absorption and allows significant portion of light to pass through in the visible range, while metallic materials such as silver (Ag) or gold (Au) have high absorption and are usually opaque in optical frequency. Note that we also include semiconductor materials into the dielectric category. Identifying the correct material selections at each layer can effectively determine how light is propagating inside the layers as well as light refraction at each interface, and thus, the optical responses in layered system. However, this is not a trivial task as materials are usually dispersive, i.e. the refractive index is dependent on the incident light wavelength. The last important design parameter is the thickness at each layer. The major usage of thickness is to introduce desirable propagation phase while light passes through each layered medium. In some special cases, this can create constructive or destructive interference phenomenon. One example is the Distributed Bragg Reflector (DBR)[43–45] that consists of alternating materials with high/low refractive index profile. When each layer's thickness is around one quarter of the effective wavelength, strong interference in the reflection side can lead to a perfect reflector that is close to 100%.

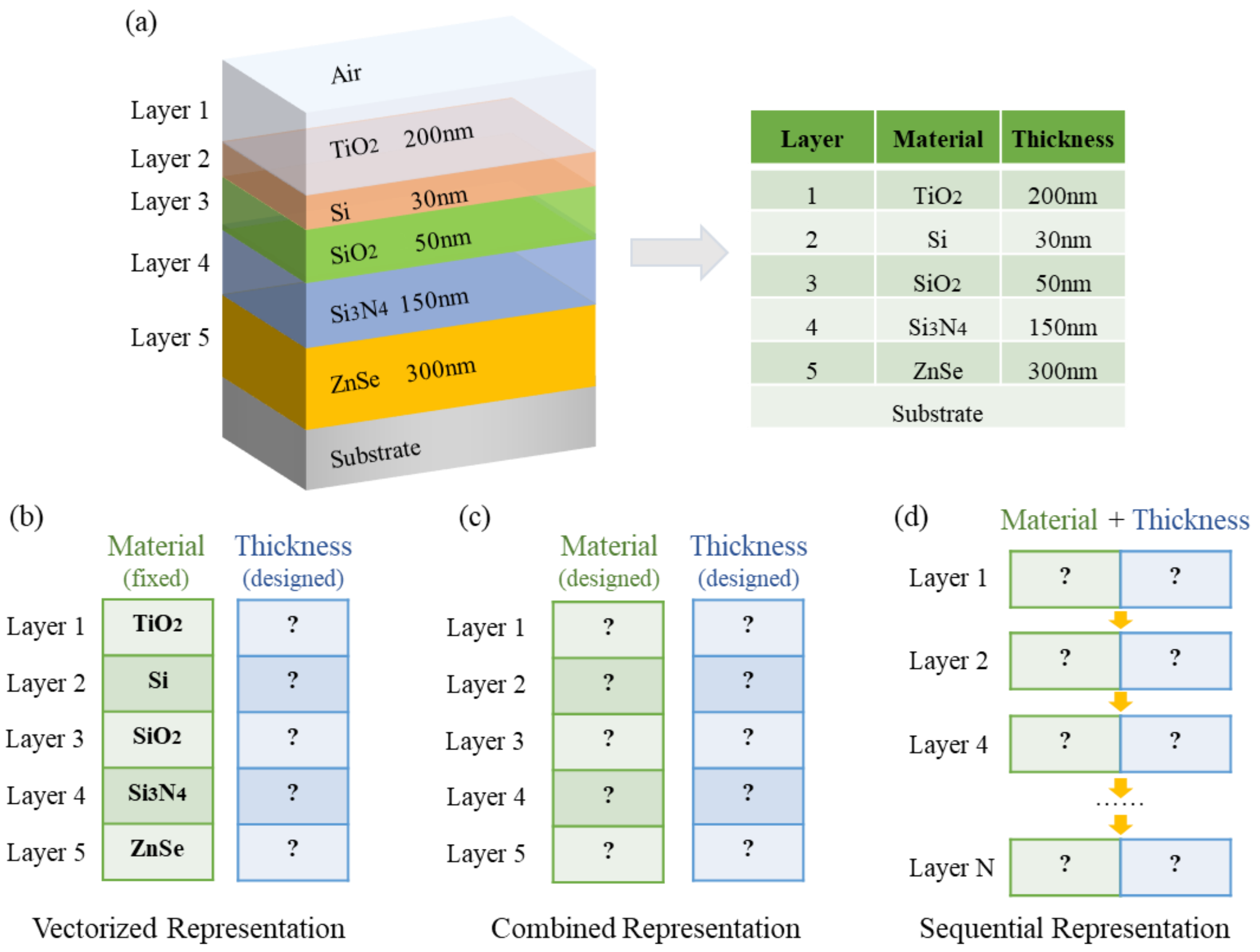

**Figure 1. Visualization of multilayer structures as well as three common types of representation.**
(a) An example of a five-layer thin film structure (left) as well as the three design parameters presented inside a table(right), including the total number of layers, the material combinations, and their thickness at each layer.
(b-d) shows three widely-used representation for a multilayer structure in inverse design algorithms. (b) is the vectorized representation, where material and total number of layers are fixed and only continuous thickness will be designed. (c) is the combined representation, where total number of layers are fixed and we need to design both material and thickness. Usually, material is a discrete variable that is selected from a pre-defined material database. (d) is the sequential representation, where all three parameters will be designed in a layer-by-layer way.

As an example, in **Figure 1** (a), we illustrate a simple five-layer structure (left side) and use a table to illustrate these three design parameters (right side). This structure is made on a glass substrate which is not treated as a design parameter. As a notation, we count the number of layers from top to bottom, i.e., the first layer is closer to the air and the fifth layer is closer to the substrate. We will use this notation for the rest of this review. Ideally, we hope that an inverse design algorithm can identify all three design parameters based on the provided target optical responses. However, there are significant difficulties when trying to incorporate them into a computer-based inverse design algorithm. For example, how can we let an algorithm to understand the number of layers, which is supposed to be a varying number? In addition, how to represent different materials in the algorithm? Remember that materials usually have dispersion, which hinders us from using a single refractive index to distinguish them.

Perhaps the easiest design parameter to deal with is the thickness, which is a continuous variable that aligns well with computational considerations. Currently, there have been three different ways of representing a multilayer structure inside the inverse design algorithm and the first type is the **vectorized representation** that only relates to the thickness parameter. We illustrate this idea in **Figure 1** (b). Within this representation, one needs to first determine how many layers to be used as well as material choices at each layer based on their prior knowledge or experience, and only use the algorithm to find out the thickness. Common algorithms include some optimization-based methods such as Particle Swarm Optimization[17] (PSO), and many machine-learning based methods such as tandem networks[19].

The major limitation for this vectorized representation comes from the fact that its design performance is heavily determined by the pre-selected materials. Considering the large variety of materials in nature, it is possible that these selected materials may not lead to a good performance for any thickness choices[9], which can leave the design algorithm running in vain. An inverse design algorithm that can automatically determine the material at each layer would be preferrable. This leads to the second type of **combined representation** (see **Figure 1** (c)). Here, the material is represented by a discrete variable and thickness is a continuous variable. Usually, there are several candidate materials to be selected and they are denoted by different integers. For example, when there are two materials, we can use 0 to represent one material and 1 for the other material. In some special cases when all candidate materials have low dispersion, a single refractive index value can be used to represent the material. Many global optimization methods use this type of representation, such as genetic algorithm[18] and memetic algorithm[46].

This combined representation has been popular for a long time. Only very recently, with the development of sequential machine learning models such as Recurrent Neural Networks[47,48] (RNN) and transformers[49], researchers start to establish the third idea of **sequential representation**, describing a multilayer structure with a sequence (see **Figure 1** (d)). Instead of splitting material and thickness separately, this sequential representation treats both material and thickness on an equal footing, and assembles multilayer structure in a layer-by-layer manner, just like in actual fabrication. Two recent models, the sequential decision network called OMLPPO[23,40], and the conditional sequence generation network called OptoGPT[39,50], are adapting this representation. Both methods have been demonstrated to automatically determine the optimal number of layers not known beforehand and terminate the design process when a target is reached.

**Optical Responses**

In the previous section we mainly talk about the output of the inverse design, namely, the design parameters. Here, we would like to briefly summarize the input of inverse design: the targeted optical responses for the designed structure. In optical frequency, multilayer structures have been widely used as transmissive/reflective filters, structural color coatings, absorber, radiative cooling devices, etc. These applications can be attributed to the following types of optical responses:

- *Transmission or Reflection Spectrum:* they are mostly used as design targets for many special filters. For example, band-pass/band-notch filter requires a transmission window/block region at the desired frequency. Designing DBR is looking for a near 100% reflection spectrum within expected wavelength region. On the other hand, a near-zero reflection spectrum is needed for designing anti-reflection coatings.
- *Absorption Spectrum*: Light absorptance is very important for many energy-harvesting and detection applications. In solar-thermal harvesting devices, a high absorption over a broadband wavelength that covers the whole solar spectrum is always desirable. Many detection devices such as spectrometer require a perfect absorption spectrum to completely remove potential crosstalk and improve the measurement accuracy.
- *Structural Color*: Compared to traditional color, structural color has unique advantages of higher spatial resolution, better stability, more environmental friendliness, etc. They can be well described using a color space such as RGB, xyY, or Lab values.

There are also many other important optical responses not listed here, e.g. combined spectrum and angular responses, which presents higher level difficulties. Different from multilayer structures that requires special representations in design algorithm, these optical responses (i.e. spectra) are usually continuous variables and can be easily represented as a high-dimension vector when integrated into design algorithm. Therefore, we will not provide more description on this.

**Challenges**

An ideal inverse design algorithm should be able to quickly output multilayer structures given any type of responses as the input. However, this can lead to significant difficulties and challenges. The first challenge is how to design the multilayer structure globally (**Global Design**). By global we mean the algorithm can automatically design for all three set of design parameters, including the total number of layers, the material sequence, and their thickness at each layer. As a comparison, local design only deals with thickness design, where the performance is heavily constrained by the fixed material selections. Global design is non-trivial as the design space is extremely large. For example, considering a five-layer structure where each layer has ten different material choices and the thickness also only has ten different discrete values, the design space can expand to $10^{10}$ different combinations of structures. It is impossible to search all potential designs exhaustively, therefore, advanced algorithms are required to navigate through this large design space and identify the optimal structure.

The second challenge is how to finish a design quickly (**Efficient Design**). Traditional optimization algorithms are based on iterative evaluations and use fitness function to update structures to reach user-defined goals. Although many numerical methods such as TMM can finish simulating a multilayer structure and obtaining its optical characteristics within one second, the iterative evaluation process often requires thousands or even more simulations, resulting in a significantly long processing time. The situation can be worse when multiple design targets are required. Speeding up the design process and making it more efficient would benefit everyone in this field.

In addition to these two major challenges related to algorithm itself, researchers and engineers are also interested in leveraging inverse design algorithm to assist practical fabrication process. Based on this, the third challenge is how to obtain multiple different designed structures with minimal efforts so that researchers can select the best one for their fabrication needs based on the availability of materials and deposition methods (**Diverse Design**). In principle, this is feasible because of the one-to-many mapping feature in the inverse design, meaning multiple structures can lead to similar optical properties. However, most existing algorithms cannot tackle this challenge because they can only output one definite design. An algorithm with probabilistic output can potentially solve this difficulty. The fourth challenge is how to integrate practical constraints into design process (**Flexible Design**). With this challenge solved, researchers may restrict the material selections and thickness range at any desired layer based on their fabrication or design needs. For example, some may require the top layer to be dielectric material to protect the next metallic layers from oxidation. The latter two challenges have not yet been extensively explored.

In this review, we will focus on these four challenges and summarize how researchers came up with innovative ideas and advanced algorithms to tackle each of them. Decades ago, some global optimizations, including genetic algorithms[18] and memetic algorithms[46], have been proposed to simultaneously design materials and thickness from a global perspective. Later around 2017, with the great success of machine learning, many different neural networks-based methods have been utilized to speed up the inverse design process. However, both methods have their own disadvantages: optimization-based algorithms are not efficient, while deep-learning based algorithms cannot design structures globally. Therefore, later around 2020, the combination of optimization and deep learning provides a solution to alleviate these problems to design in the global space while still making the algorithms fast. However, the iterative process still takes quite a long time. Finally, in recent two or three years, with the development of sequential learning, researchers start to treat the multilayer structure as a sequence and eventually developed OptoGPT that can solve all four challenges, declaring victory of the long-lasting design issue in multilayer structure. In *Table 1*, we summarize all these impactful algorithms and list their accomplishments.

**Table 1. A comparison of different inverse design algorithm for multilayer structure.**

| Methods | Reference | Year | Multilayer Representation | Global Design? | Efficient Design? | Diverse Design? | Flexible Design? |
|---|---|---|---|---|---|---|---|
| Needle Optimization | [16] | 2007 | Combined | Yes | No | Yes | No |
| Genetic Algorithm | [18] | 2008 | Combined | Yes | No | No | No |
| PSO | [17] | 2014 | Vectorized | No | No | No | No |
| Memetic Algorithm | [46] | 2018 | Combined | Yes | No | No | No |
| Tandem Networks | [19] | 2018 | Vectorized | No | Yes | No | No |
| MDN | [20] | 2020 | Vectorized | No | Yes | Yes | No |
| VAE | [26] | 2021 | Vectorized | No | Yes | No | No |
| GAN | [51] | 2022 | Vectorized | No | Yes | Yes | No |
| GLOnet | [52] | 2021 | Combined | Yes | No | Yes | No |
| NEUTRON | [53] | 2022 | Combined | Yes | No | Yes | No |
| OMLPPO | [23] | 2021 | Sequential | Yes | No | Yes | No |
| OptoGPT | [39] | 2024 | Sequential | Yes | Yes | Yes | Yes |

## Optimization-based methods

Since decades ago, heuristic optimization-based methods have become essential tools in tackling inverse design challenges, offering robust and flexible approaches to explore the vast solution space and find near-optimal designs. Although dealing with inverse design, these methods are mainly based on iterative forward simulation and follow a similar pipeline shown in **Figure 2**. When given a design target, the algorithm will first start from a random structure and obtain its simulated optical responses. Then, the difference between the simulated responses and target responses will be evaluated and used to calculate the fitness function. If the fitness function does not meet the termination criteria, the algorithm will update the initial structure and keep running this iterative process until reaching satisfaction. After sufficient iterations, a structure with desired optical responses can be found. Different optimization methods, including needle optimization, particle swarm optimization, genetic algorithms, differ from each other in the way how they update the structure in each iteration, with illustrations can be found in **Figure** 2. A summary and benchmark of optimization-based methods on photonic applications can be found in Ref [30,38].

Optimization-based methods mainly deal with thickness or material design. At the beginning, researchers use the vectorized representation to only optimize the thickness. One of such examples is Particle swarm optimization (PSO), an evolutionary algorithm inspired by the social behavior of bird. PSO maintains a set of particles that move through the solution space and are influenced by their own best-known position and the best-known positions of the entire particles. In this case, each particle is a vector of layered thickness that can adjust its trajectories based on its own experience and the experience from neighboring particles, balancing the exploration and exploitation. Rabady[17] *et al.* used PSO to design band-pass filters and demonstrated good performance (see **Figure 3** (a)). PSO has also been applied in various real-case multilayer thin film applications, including transmissive spectrum design[54], inverse extraction of thickness and optical constants[55], transmissive filters for image differentiation[56], etc.

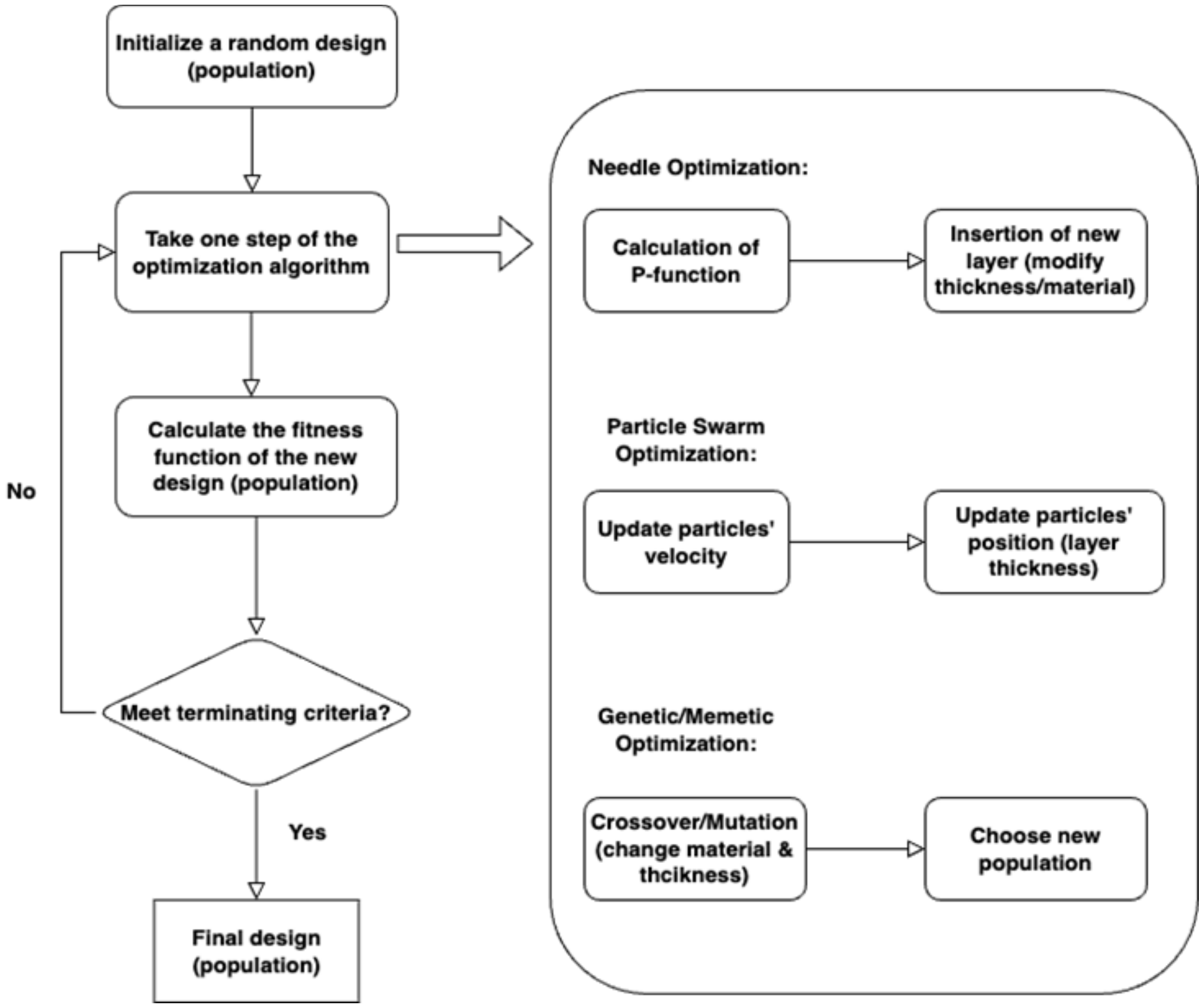

**Figure 2. A general pipeline of optimization-based methods for inverse design.**

To further explore global design of both material and thickness, Tikhonravov[15] *et al.* introduced needle optimization for optical coating design. Needle Optimization iteratively adds or adjusts layers in a thin film design to improve its optical properties toward the design target by modifying their thickness or refractive index of the material configuration (see **Figure 3** (a)). The needle optimization evolves gradually, allowing for the generation of multiple solutions with nearly the same performance metrics, providing the designers with a range of options to select the most practical and manufacturable design. Another example is the application of genetic algorithms (GAs) to the design of optical multilayer structures by Martin[18] *et al.*. Genetic algorithms are stochastic optimization methods inspired by the principles of natural selection and genetics. They operate by generating an initial population of potential solutions and iteratively evolving these solutions through selection, crossover, and mutation operations to maximize a fitness function, which is usually taken as the merit function commonly used in traditional approaches. In each iteration, the individuals with the highest fitness score are more likely to be selected for reproduction, ensuring that their genetic material is passed on to the next generation. Crossover combines parts of two parent solutions to create offspring with potentially better performance, while mutation introduces random changes to individual solutions to maintain diversity within the population and prevent premature convergence to local optima. The authors demonstrated the effectiveness of this algorithm by designing the refractive index and thickness of each structure layer for three different optical filters: antireflection coating, beam splitter, and rejection filter (see **Figure 3** (c)). This idea has also been applied to the design of broadband reflectors[57], edge filters[58], etc.

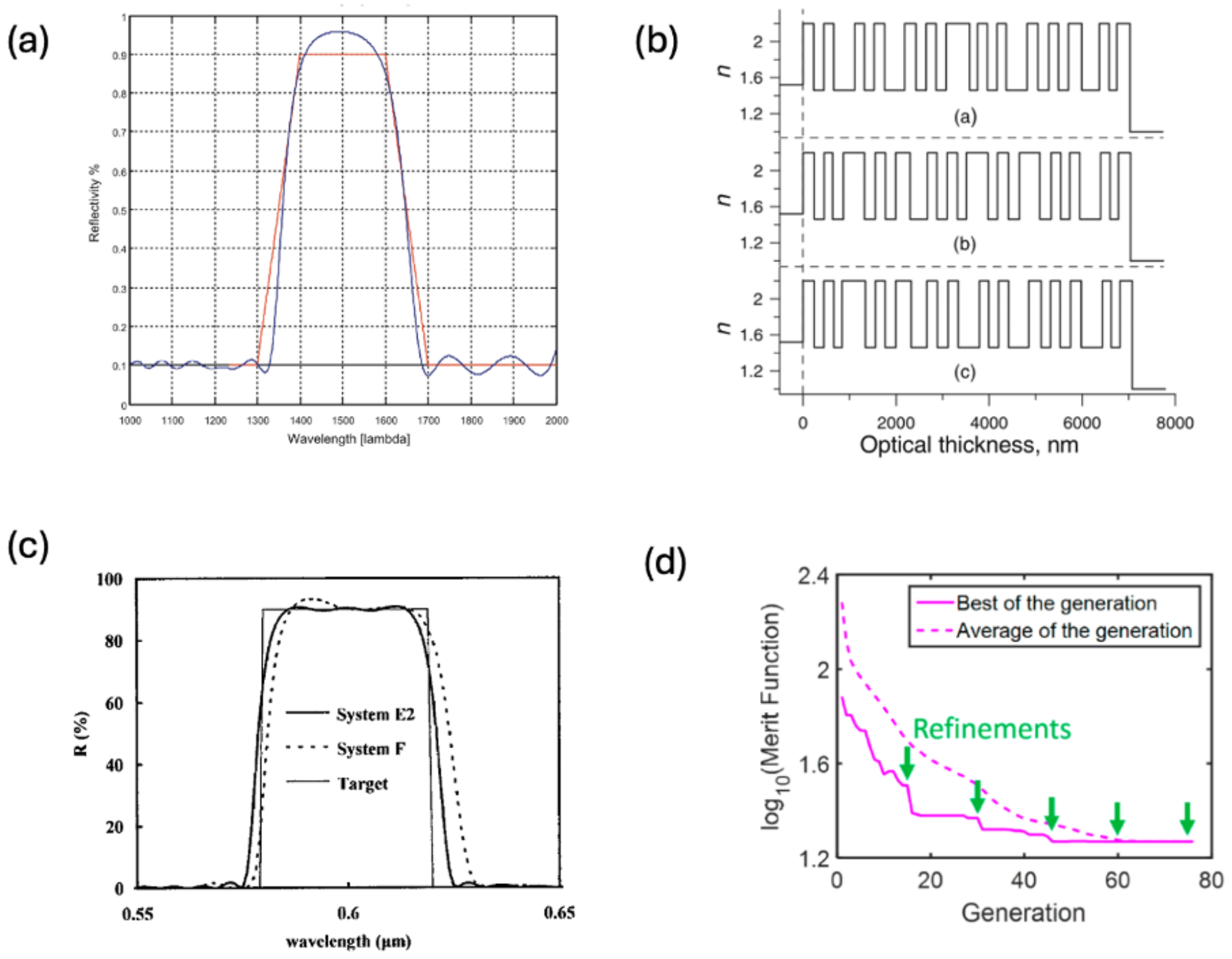

**Figure 3. Several design examples for optimization-based methods.**
(a) Design performance of Particle swarm optimization (PSO) algorithm for a band-pass optical interference filter in the range of 1000–2000 nm wavelengths. Reprinted from Ref. [17]. Copyright @ 2014, Elsevier.
(b) A demonstration of versatile design using Needle Optimization. Reprinted from Ref. [16]. Copyright @ 2007, Optica Publishing Group.
(c) Genetic Algorithm designing a rejection filter of 90% over the 0.55-0.65 micrometer region. Reprinted from Ref. [18]. Copyright @ 2008, Optica Publishing Group.
(d) Iterative refinement of memetic algorithm with mixed-integer programming (MIP) to design better generations. Reprinted from Ref. [46]. Copyright @ 2018, ACS Publications.

However, using refractive index to represent materials restricts the design target to be narrow band because of materials' dispersion. To design for a broader wavelength, Shi[46] *et al.* introduced a mixed-integer programming (MIP) with a genetic algorithm solution to optimize the design of multilayer thin films. MIP can handle both discrete and continuous variables in the optimization process. In this context, the discrete variables represent different dielectric materials, while continuous variables correspond to the thickness of each material layer. This algorithm improves upon traditional genetic algorithm by introducing discrete materials into the crossover and mutation steps (see **Figure** ***3*** (c)). The authors demonstrated state-of-the-art performance using the algorithm on two design tasks with a broad wavelength range: radiative cooling devices and incandescent light bulb filters. The GA with MIP algorithm's ability to design a variety of solutions with distinct material and thickness permutations discovered innovative designs for multilayer thin films at a global solution space for wide wavelengths, materials, and thicknesses.

## Deep Learning Methods with Vectorized Representation

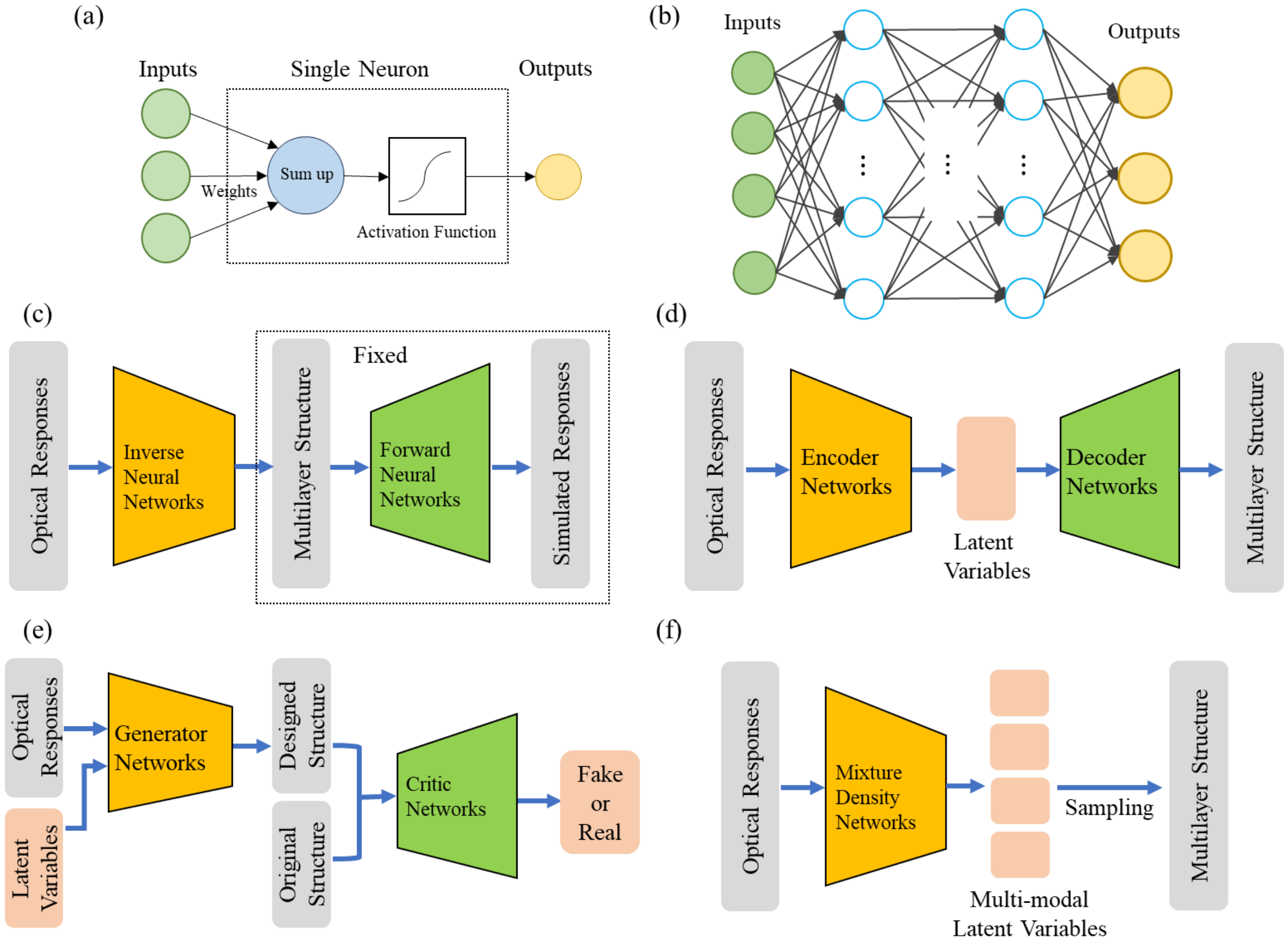

**Figure 4. Illustration of different types of neural networks.**
Visualizations of (a) a single neuron, (b) Multilayer-Perceptron (MLP), (c) tandem networks, (d) Variational Auto-Encoders (VAE), (e) Generative Adversarial Networks (GAN), and (f) Mixture Density Networks (MDN).

The fundamental unit inside neural networks is the artificial neuron. As shown in **Figure 4** (a), a single neuron first takes one or multiple inputs, then sums up all these inputs multiplied by learnable weights and outputs results through a nonlinear activation function. In this way, information can be passed in, get processed and propagated to the next level. Usually, a single neural network (NN) can contain thousands or millions of neurons, leading to millions or billions of learnable weights. After training on a large dataset, these learnable weights are updated to capture the general mapping between the inputs and outputs and process extremely complicated information such as image classification[59–61], speech recognition[62–64], sequence decision[65–67], language understanding[49,68–70], etc. Different types of neural networks differ in the way how the neurons are connected, and therefore, suitable for processing different types of data. For example, Convolutional Neural Networks (CNN)[59] use the convolutional layers to capture spatial correlations, thus widely used for processing and analyzing image-type data. Recurrent Neural Networks (RNN)[47] employ recurrent connections to obtain temporal dependencies, making them well-suited for sequential data.

In previous discussion, we conclude that most of these optical responses (inputs) can be denoted as vectors while multilayer structures (outputs) can be denoted using either vectorized, combined, or sequential representations. However, combined representations require a discretized output for material design, which is non-trivial for NN since their outputs are continuous variables. Aso, it was not until recent two or three years that researchers have figured out the sequential representation. Therefore, most of early works that used deep learning for inverse design seek to solve the vector-to-vector mapping. A general working process is to first generate a training datasets containing multilayer structures and corresponding optical responses using simulation (e.g., TMM), then train the NN model to fit the inverse mapping function from optical responses to multilayer structures. Once trained, the model can be used to design multilayer structures for different optical responses instantaneously, providing a much faster alternate than optimizations, which usually takes minutes or hours.

In terms of the deep learning models for vector-to-vector mapping, perhaps the most straightforward way is the Multilayer-Layer Perceptron (MLP), where neurons are stacked layer-by-layer and all neurons are fully connected to other neurons in nearby layers (see **Figure 4** (b)). However, directly using MLP to inverse design will give inaccurate results. Because of the one-to-many mapping issue, i.e., there are multiple possible structures for a given optical response, it is difficult to minizine the loss function (the difference between model outputs and target outputs) and have NN model converged during training. Eventually, the model will learn to output an averaged structure, which usually do not have the desired optical responses[34]. Therefore, special buildup of neural networks is required to handle this one-to-many mapping issue. Many different types of neural networks have been proposed and demonstrated effectiveness in inverse design. Based on the type of outputs, they can be classified into two different categories: 1) deterministic NN, mainly including tandem networks, and 2) stochastic NN (or generative NN), including Generative Adversarial Networks (GAN), Variational Auto-Encoders (VAE), and Mixture Density Networks (MDN). In the remining section, we are going to discuss each model and summarize their variants and applications.

**Deterministic NN**

In deterministic NN, the model always maps the input into a single output. The most popular one is the tandem networks. As the name "tandem" suggests, tandem networks connect two NN in series (see **Figure 4** (c)): the Forward Neural Networks (FNN) and the Inverse Neural Networks (INN). The FNN take in the structural parameters and output the predictions of their corresponding optical responses. Once trained, we can use FNN to quickly predict the optical responses for a given structure with high accuracy. On the other hand, INN take in the target optical responses and output predicted structures. The training of tandem networks involves a two-step training procedure for each individual network. In the first step, we need to train FNN separately to obtain a fast surrogate predictor for optical responses. The second step is to connect the output of INN to this pre-trained FNN and use FNN to supervise the learning of INN. In this way, we can resolve the one-to-many mapping issue by only predicting one structure suggested by FNN.

Tandem networks is first introduced in Ref.[19] to design transmission spectra for multilayer structures. The multilayer system in consideration is a twenty-layer structure with $SiO_2/Si_3N_4$ alternating material combinations. Using this method, the authors have demonstrated excellent design performance for arbitrary spectra, as well as the special Gaussian shaped spectra. In addition, it only takes a fraction of a second to finish a design, which is much faster than optimization-based methods. Later, extensive works have dedicated to utilizing tandem networks to design different multilayer structures and tackle various design problems efficiently, including designing Fabry-Perot-cavity-based color filter[71], organic light-emitting diode[72], etc. In addition to multilayer structure, tandem networks are also widely used in other types of photonic structures and target responses. For example, Gao[73] *et al.* used tandem networks to design reflective structural color for silicon nanorod metasurface; Ma[74] el.al combined tandem networks to design reflection spectrum for three-dimensional chiral metamaterials; He[75] *et al.* utilized tandem networks to design far-field radiation profile for plasmonic nanoparticles.

There are also many variants to improve the design performance of tandem networks. For example, Xu[24] *et al.* modified and benchmarked different activation functions and loss functions used in tandem networks to find the optimal schema. They found that using ReLU activation function[76] together with intermediate structural loss can significantly increase the design accuracy. On the other hand, many researchers seek to find alternate neural networks to replace the MLP architecture used in INN and FNN. For example, Chen[22] *et al.* proposed to use transformer architecture for both INN and FNN by leveraging the self-attention mechanism to effectively extract and process information. In this work, they proposed meta-material spectrum transformer (MST) to design a high-performance broadband solar metamaterial absorber that exhibits high average absorptance of 94% in a broad solar spectrum. In addition, considering that the original tandem network can only output one designed structure, Yuan[25] *et al.* came up with the idea of connecting the INN to multiple different FNN, which they called multi-head tandem networks, to expand its capability of outputting multiple structures.

**Stochastic NN**

Different from deterministic NN, stochastic NN (also called generative NN) can output multiple different predictions given the same input. This is because these models take in one latent variable randomly sampled from a normal distribution as an extra input, which enables them to generate different outputs conditioned on this extra input. Therefore, they can intrinsically solve the one-to-many mapping issue. There are three major types of generative NN that are widely used in inverse design: VAE[77], GAN[78], and MDN[79]. We illustrate their model architectures in **Figure 4** (d-f), respectively. There are already many reviews[32,33,35] and benchmark[28,80–82] that discuss these models, but towards a more general audience on the photonic inverse designs. Here, we mainly summarize the work done for multilayer structure designs.

**Figure 4** (d) shows the model architecture for VAE, which originates from information theory: it first uses a NN (called encoders) to encode the high-dimensional input information into a low-dimension hidden representation, usually with a normal distribution, then uses another NN (decoders) to decode hidden representations back into the corresponding output. During inverse design, each input (target optical responses) will be encoded to different normal distributions with different means and variance. Sampling from this distribution and sending it through the decoder will give the designed structure. Since this sampling process is random, VAE can output multiple structures when decoding from different latent variables, thus solving the one-to-many mapping issue naturally. For example, Zandehshahvar[26] *et al.* used the VAE to design multilayer structures composed of consecutive layers of $SiO_2$ and $TiO_2$. In their work, they demonstrated that when given a single transmission spectrum as input, their model can output three different structures (**Figure 5** (a) left side). The simulation shows that their simulated spectra are close to the target spectrum (**Figure 5** (a) right side). VAE have also demonstrated successful applications in other types of photonic structures, including Fano-type resonant nanopillars[83], hybrid plasmonic/phase-change material metasurface[84], periodic array of Au nanoribbons[85], etc.

Another type of stochastic NN is GAN (**Figure 4** (e)). It also contains two neural networks: the generator NN that generate designed structures based on the random latent variables and the input optical responses, and the critic NN that learn to distinguish if the input structure comes from model or not. Different from VAE, GAN is originated from game theory and these two NN learn to generate by combating with each other: the generator tries to generate outputs that are distributed as close to the training dataset as possible to fool the critic, while critic tries to distinguish the generated outputs from the ground truth in training dataset. A successful application is demonstrated in Dai's work[51], where they used GAN to design a transmissive Fabry-Perot-cavity-based color filter using the three-layer Ag-$SiO_2$-Ag structure and identified an average number of 3.58 solutions for each color. Experimental results in **Figure 5** (b) also validated the effectiveness of these designs.

MDN is the third type of stochastic NN (see **Figure 4** (f)). Different from VAE and GAN which utilize single-modal normal distribution, MDN models the design parameters as a multimodal probability distribution. Here, each distribution in this multi-modality is parameterized by a different normal distribution with unique mean, variance, and weight. The total number of distributions is a hyperparameter that is usually pre-determined, and can be tens or hundreds. After training, the designed structure will be obtained after sampling from this predicted multimodal distribution, which inherently solve the one-to-many mapping issue. As an example, Unni[20] *et al.* utilized MDNs with 16 mixtures to design 10 layers of alternating $SiO_2$ and $TiO_2$ structure. Their results in **Figure 5** (c) demonstrated that the MDN can give multiple structures that satisfy the desired transmission spectrum.

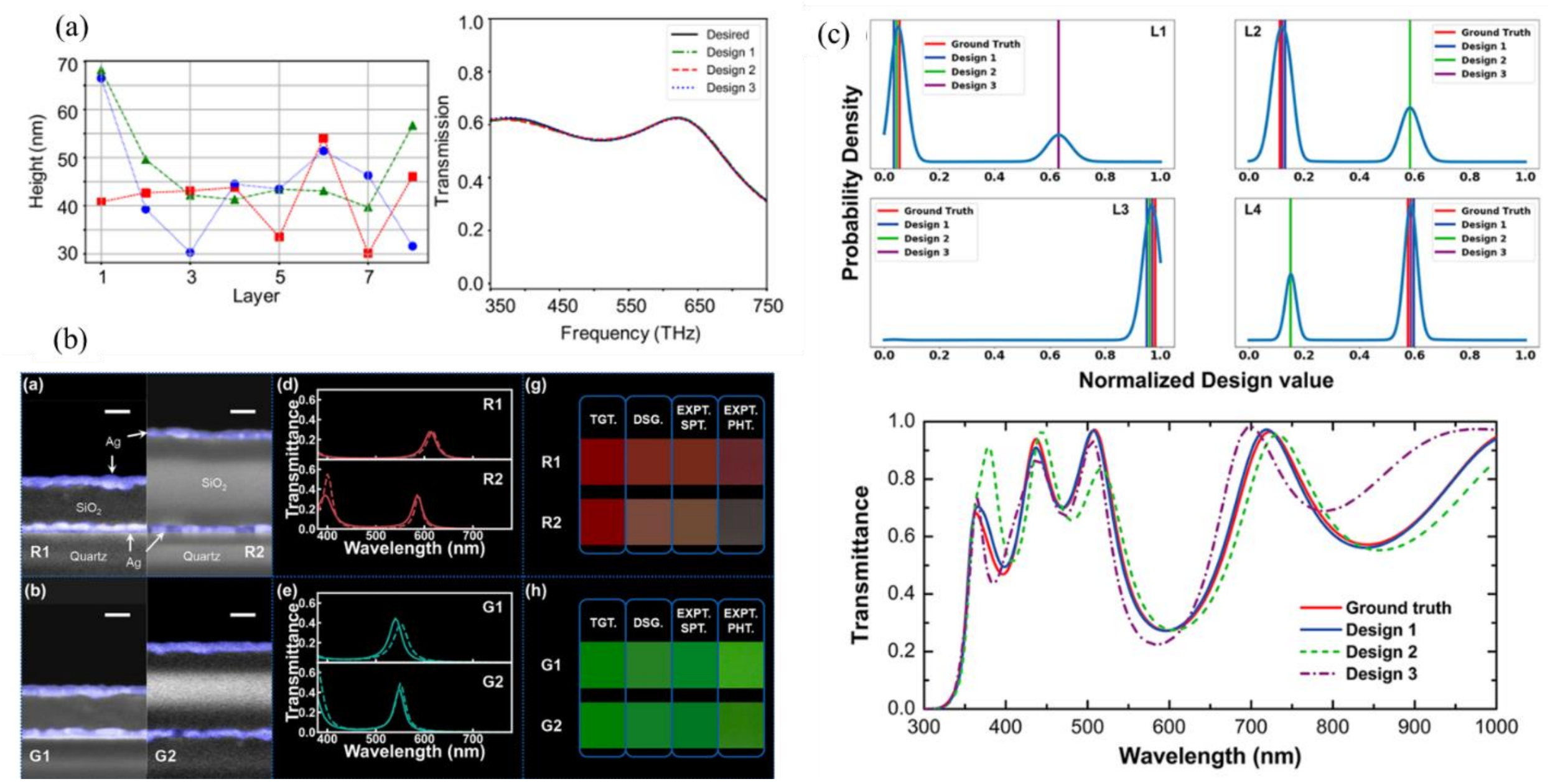

**Figure 5. Stochastic NN for inverse design with multiple outputs.**
(a) VAE can output multiple designs that satisfy the desired transmission spectrum. Reprinted from Ref. [26]. Copyright @ 2021, Optica Publishing Group.
(b) A three-layer structural color designed by GAN showing different structures. Experimental results also verified the design performance. Reprinted from Ref. [51]. Copyright @ 2022, Elsevier.
(c) The output probability from MDN showing a multi-modality distribution which corresponds to different designs. All designs show desired performance in the transmission spectrum. Reprinted from Ref. [20]. Copyright @ 2020, ACS Publications.

# Optimization-combined Deep learning

Although optimization and deep-learning based methods are widely used for inverse design, both have their own strengths and limitations. Optimization has much better flexibility to incorporate the material into design consideration. Designing in this global space with both materials and thicknesses usually leads to better performance than deep learning-based methods. However, optimization requires iterative evaluations to minimize the fitness function, which can be time-consuming. In addition, one needs to restart the optimization process from scratch whenever encountering a new design target, making them task-specific. On the other hand, because deep-learning based methods usually use neural networks to learn a general mapping from the space of optical responses to the space of multilayer structure, they can finish a design task much faster than optimization and can quickly adapt to a different design target. The probability output for stochastic NN also enables multiple designs to solve one-to-many mapping issue. However, these neural networks usually have vectorized outputs, making it difficult to incorporate the discrete materials into outputs, thus, resulting in limited design performance. A natural question to ask is if it is possible to combine both methods together and obtain greater benefits. It is encouraging to see that many researchers have already started to do so. Based on how these two methods are integrated together, we can classify them into three categories and we will summarize each of them below.

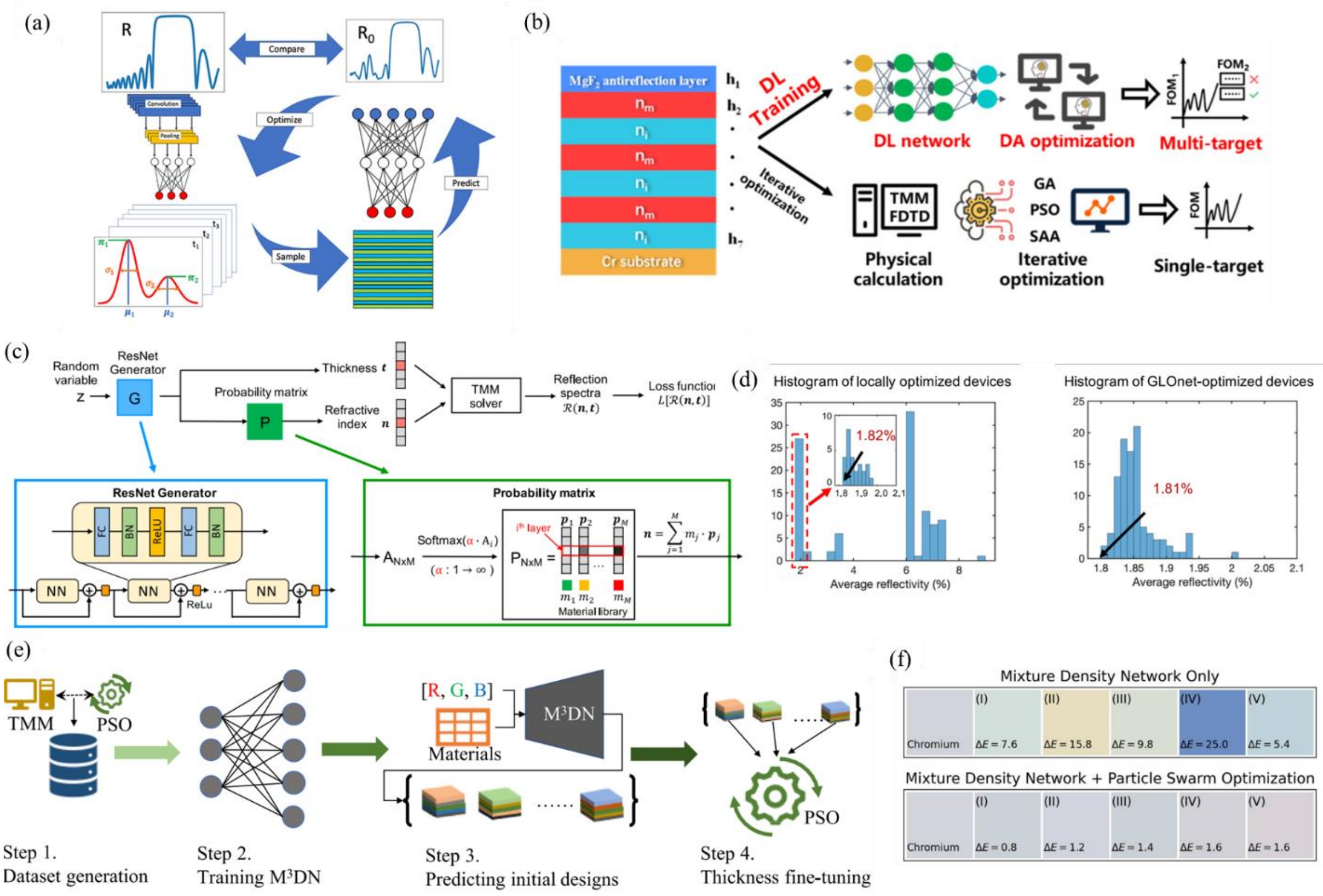

**Figure 6. Multiple ways to combine optimization and deep learning together.**
(a) A mixture-density-based tandem optimization network. Reprinted from Ref. [21]. Copyright @ 2021, Elsevier.
(b) Deep learning combined with multi-objective double annealing algorithms. Reprinted from Ref. [86]. Copyright @ 2023, Elsevier.
(c) Global optimization networks (GLOnets) and (d) the design performance comparison using GLOnets and traditional optimization. Reprinted from Ref. [52]. Copyright @ 2021, Elsevier.
(e) NEUTRON: Neural particle swarm optimization for material-aware inverse deign and its design performance combined to NN only (f). Reprinted from Ref. [53]. Copyright @ 2022, Cell Press.

## Neural Networks as a fast surrogate simulator

Neural networks are not only fast for inverse design, but also a fast surrogate simulator to predict the optical responses given the structure[19,50,87]. Considering that optimization requires an iterative evaluation process, it is straightforward to use NN as the surrogate simulator to speed up the optimization process. For example, Unni[21] *et al.* proposed a mixture-density-based tandem optimization network to design high reflectors. As shown in **Figure 6** (a), they first pre-trained a forward simulator network to instantly evaluate any designs, then run optimization algorithms to quickly identify the optimal structure. To further speed up the design process, they trained a mixture-density network to give some initial starting structures for optimization. Using this method, they have successfully designed an ultra-broad high reflector using 20-layer structure. In another work, Ma[86] *et al.* combined deep learning and multi-objective double annealing algorithms to simultaneously maximize solar spectrum absorption and minimize infrared radiation. As shown in **Figure 6** (b), they first trained a MLP for fast evaluation, then use the numerical optimization to find the multilayer structure that leads to the best performance. Compared to traditional optimization, this deep learning-aided optimization can drastically reduce simulation time and make optimization several orders of magnitude faster.

**Neural Network as surrogate gradient update**

In addition to being fast, another advantage of neural networks is that they are differentiable. After learning from the dataset, NN can build up a continuous and analytical mapping between input and output spaces, even if the relationship in original data is too complicated to describe with closed-form equations. Therefore, we can use NN to calculate the gradient information and update the input structures directly. Motivated by this idea, Fouchier[88] *et al.* first trained a forward simulator, then fix the parameter in NN and use backpropagation to calculate gradient and update input structure parameters. They successfully used this method to design coatings with different light scattering properties. Another typical work is called global optimization networks (GLOnets)[52,89,90]. As shown in **Figure 6** (c), GLOnets combine generative NN with a TMM simulator to perform population-based optimization. During design process, the generative NN will output thickness information and refractive-index profile using a probability matrix. The combined multilayer structure output will go through TMM solvers and obtain a loss function, which will be used to calculate backward gradient and update weights in generative NN. After optimization, the generative NN will map the input optical responses to a designed structure. Using this combined method, they have identified an anti-reflective coating with 1.81% averaged reflection, which is lower than traditional optimization methods (See **Figure 6** (d)). In a later work, Zhang[91] *et al.* used GLOnets to design optical transfer function of a multilayer film at wavelengths of 532nm and 633nm and successfully achieved incoherent differentiation with a high resolution of 6.2um, opening the door for high-speed imaging processing, object tracking, and disease diagnosis.

**Two-step design process**

Different from previous two methods that use NN as a surrogate model, the third type splits the design process into two steps: first use NN to design material combinations, then use optimization to identify the suitable thickness. An example is the algorithm of Neural Particle Swarm Optimization (NEUTRON), which was introduced by Wang *et al.* in Ref. [53] to inverse design structural colors. The idea was illustrated in **Figure 6** (e). Here, instead of design material and thickness at the same time, NEUTRON first uses a classifier to predict the most suitable materials at each layer for a given layered structure, then uses MDN to give the initial thickness guess in a probability distribution form and combine PSO to refine and optimize the thickness and achieve optimal performance. Using this two-step process, they demonstrated a much more accurate structural color design compared to single MDN, and used it to output structures for a large number of colors in a painting. We give one design example in **Figure 6** (f).

# Deep Learning Methods with Sequential Representation

Sequence modelling and understanding is an important research field in machine learning as most of the data in our world can be treated as sequences, including natural language[49], human voice[92], time-series[93], video stream[94], etc. Different from vectorized data with pre-determined shape, sequential data usually have variable length and requires special architecture in NN, e.g., the recurrent connection inside RNN[47,48], to handle the serialized information. Recently, researchers have observed that the varying numbers of layers in a multilayer structure is analogous to the varying lengths in sequential data. Therefore, they have started to treat multilayer structure as a sequence and use sequential NN to deal with inverse design. In this way, researchers can completely automate the inverse design process for the first time: NN can design the material and thickness simultaneously at each layer, and can also design the total number of layers by smartly terminating the design process whenever NN think the current design is good enough. This points to bright future for multilayer inverse design. Currently, there are two different sequential learning methods used for inverse design: sequential decision process and conditional sequence generation.

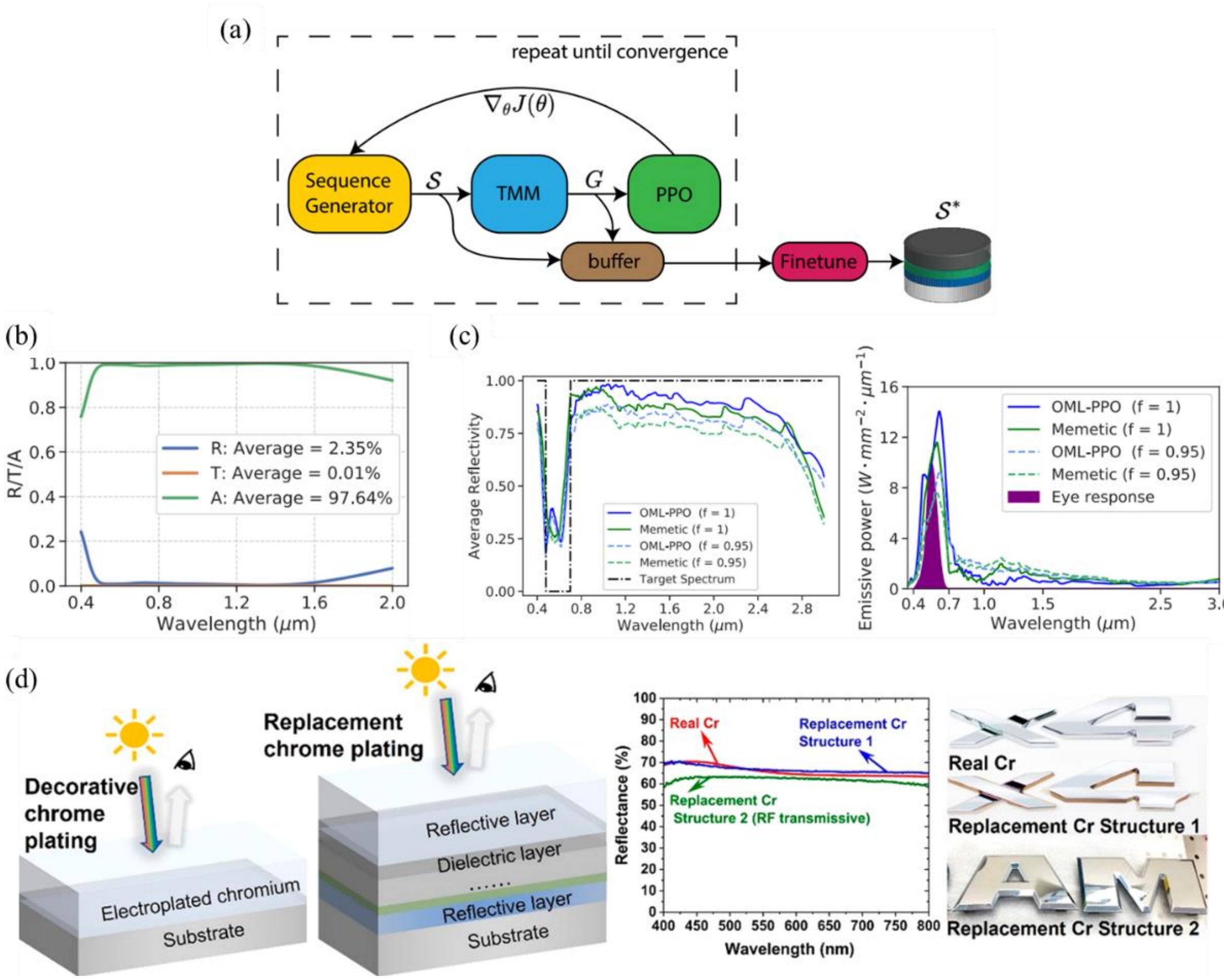

**Figure 7. Sequential decision process for multilayer design.**
(a) The working pipeline of OMLPPO algorithm.
(b) The perfect absorption spectrum designed by OMLPPO, which is 2% higher than demonstrated human designs.
(c) The incandescent light bulb filters designed by OMLPPO, which exhibits 8% performance improvement than optimization methods. (a-c) are reprinted from Ref. [23]. Copyright @ 2021, IOP Publishing.
(d) The decorative Chrome-looking reflective coating designed by OMLPPO. Reprinted from Ref. [40]. Copyright @ 2023, ACS Publications.

## Sequential Decision Process

The first type of sequential learning model is based on sequential decision process[66], where the model needs to decide what to do next based on current situation. It has been widely used in robot manipulations[95], game playing[65,67], self-driving cars[96,97], etc. In 2021, Wang[23] *et al.* proposed to use sequential decision process for multilayer design and introduced the algorithm of OMLPPO based on the reinforcement learning algorithms[66,98]. In this work, OMLPPO designs and finalizes the multilayer design in a layer-by-layer manner: after figuring out the material and thickness for the first few layers, it will decide what type of material to use for the next layer as well as its corresponding thickness. This process will keep going until the designed structure reaches the desired performance, or the sequence length (namely, the number of layers) equals the maximum allowable number of layers. The proposed OMLPPO algorithm learns to make sequential decisions in a trial-and-error way and uses iterative evaluation to update designs and approach ideal structure. The complete pipeline is shown in **Figure 7** (a). Although this iterative process looks similar to heuristic optimizations, OMLPPO can learn multilayer structure design domain knowledge during training and leverage them to improve the design performance. For example, the algorithm automatically designed a five-layer ultra-wideband absorber with 97.64% average absorption, which is ~2% than the one designed by human experts (**Figure 7** (b)). They also used the algorithm to design incandescent light bulb filters and achieved 8.5% higher visible light enhancement factor than the structure designed by optimization (**Figure 7** (c)). Because of the high performance, OMLPPO has also been adapted to solve other non-trivial design tasks. For example, Saha[40] *et al.* identified two multilayer structures that have similar visible reflection spectrum as decorative Cr, serving as an environmentally friendly replacement for the harmful and toxic chemical finishing process (**Figure 7** (d).

**Conditional Sequence Generation**

Another type of sequential learning model is conditional sequence generation, where the model needs to generate serialized output based on the input. One important example is the language modelling[99]: to predict the probability of a sequence of words conditioned on previous words and construct coherent and contextually relevant texts. It is the backbone for a wide range of natural language processing (NLP) applications, including machine translation[100], question answering[101], text generation[102], etc. The recent introduction of GPT-type models[69,102–104] which use the decoder part of the Transformer[49] model, not only improves the learning efficiency of language modeling, but also enhance the model's ability to generate more coherent texts by scaling up the model size and training dataset, leading to the development of powerful language models, including ChatGPT[105], Llama[106], BART[70], etc. The success of GPT model in language modeling also intrigues researchers to solve other sequential generation problems in scientific and engineering applications, including DNA sequence generation[107], protein prediction[108], molecule modeling[109,110], etc.

Inspired by these successful GPT models, in 2024, Ma[39] *et al.* proposed the OptoGPT model and completely tackled these four challenges for multilayer design. **Figure 8** (a) and (b) show the analogy between GPT models and the proposed OptoGPT model. For GPT models, given some language texts as input, e.g., a question or task description, they can generate a sequence of words that are related to the inputs. Similarly, given the input of optical responses (in their case, is reflection and transmission spectrum), the OptoGPT model can output the serialized representation of multilayer structures directly, without any iterative process.

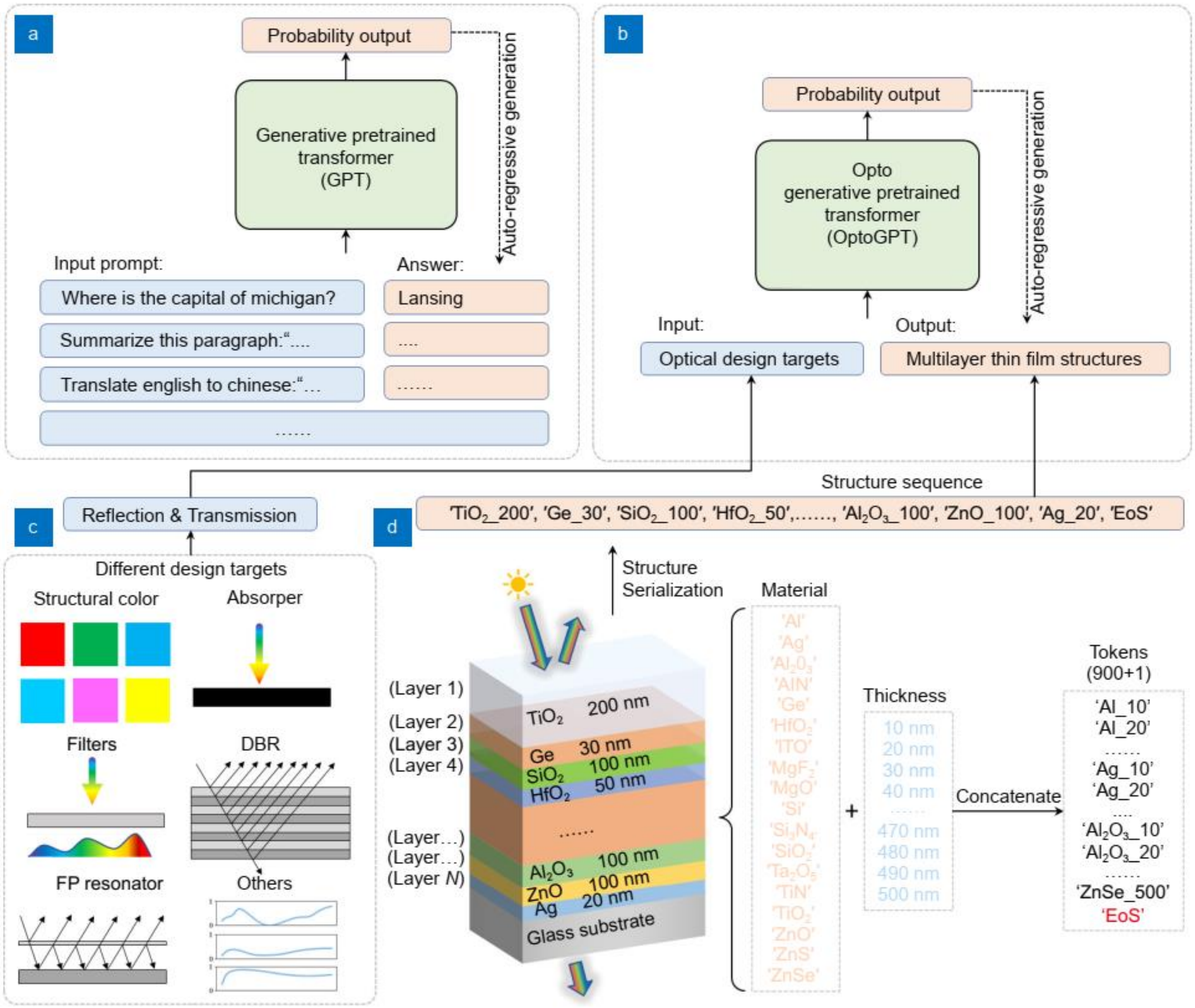

**Figure 8. Idea of conditional sequence generation using OptoGPT model.**
(a) and (b) shows the diagram of general GPT models in NLP and the OptoGPT model in multilayer inverse design, respectively.
(c) Different types of input to the OptoGPT model as the design target.
(d) One example of the "structure tokenization" and "structure serialization" for a N-layer structure on the glass substrate. All figures are reprinted from Ref. [39]. Copyright @ 2024, OE Journals.

To make OptoGPT work, we first introduce the "structure serialization" technique to convert the material and thickness information at each layer into a single token by simply concatenating them together. As shown in **Figure 8** (d), a token is a human-readable notation similar to a word or a phrase in language. Then, we proposed the "structure serialization" technique and add these tokens one by one to form the sequence representation of multilayer structure. In this way, the OptoGPT model can effectively design for both material and thickness information, and automatically determine the optimal number of layers. The input to the OptoGPT model is optical responses. To make OptoGPT well suited for different types of optical responses, we propose to design for both transmission and reflection spectrum and use multiple techniques to extend the input to diverse responses, including absorption spectrum, reflective and transmissive structural color, etc. Examples can be found in **Figure 8** (c).

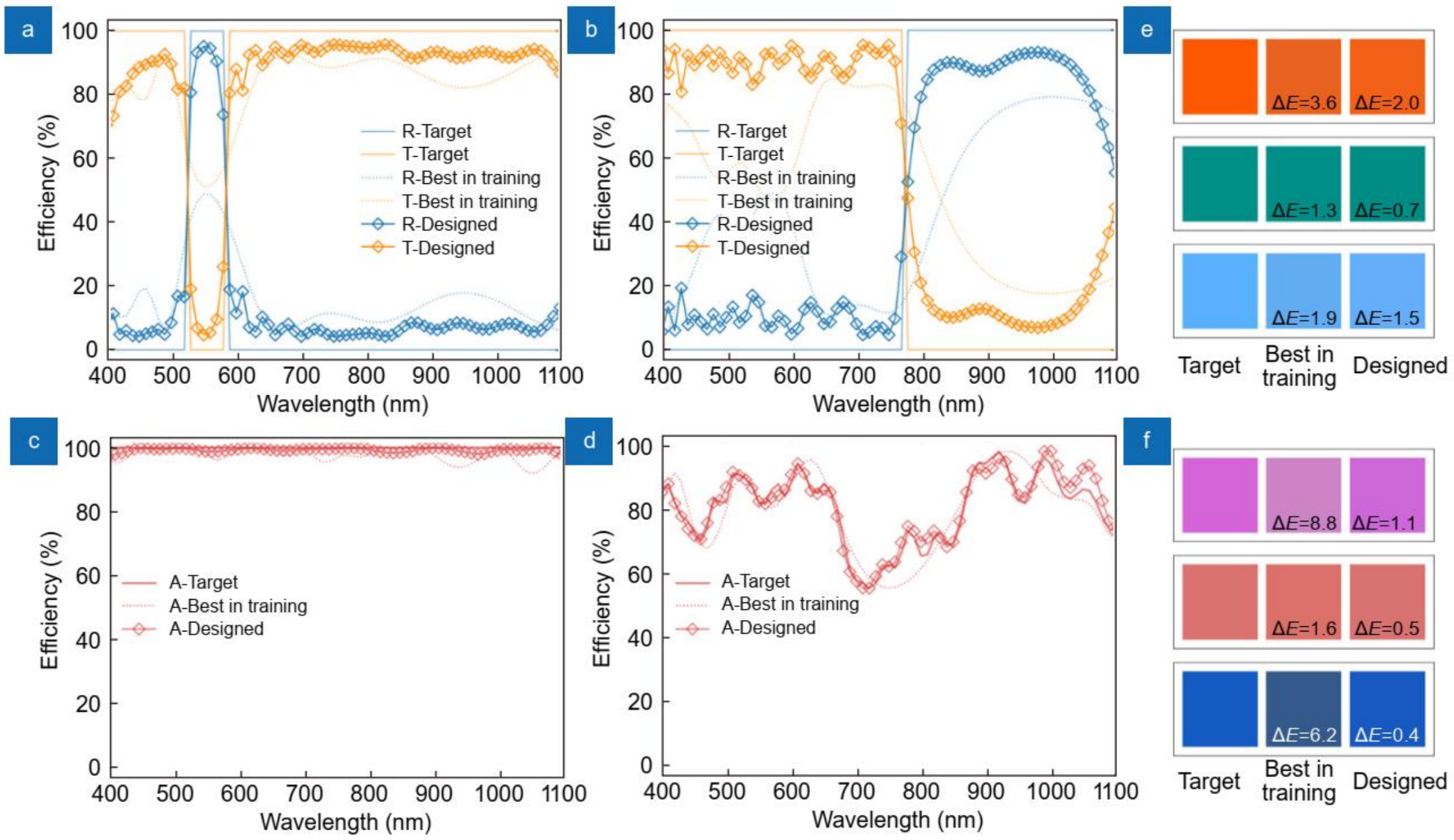

**Figure 9. Examples of different types of optical designs that OptoGPT can design with.**
(a) Design for a band-notch filter at 550nm.
(b) Design for high reflection in near-infrared.
(c) Design for perfect absorber.
(d) Design for arbitrary absorber.
(e-f) Design for reflective/transmissive type structural color. All figures are reprinted from Ref. [39]. Copyright @ 2024, OE Journals.

OptoGPT is a foundation model[111] that once trained it can be used to solve extensive downstream tasks and applications to design multilayer thin film structure. OptoGPT demonstrated a unified inverse design towards a range of applications, including transmissive/reflective filter, high reflection filter, perfect absorber, arbitrary absorber, reflective/transmissive structural color, etc. **Figure 9** gives some inverse design examples. Notably each design task can be completed within 0.1 seconds. In addition, because the output is based on probability sampling, therefore, OptoGPT can give different designs by running inference multiple times, which solves the diversity challenges. To solve the flexible design challenge, we further proposed "probability resampling" to design structures that can satisfy arbitrary constraints during fabrication or other practical requirements. Using finetuning and "mixed sampling", OptoGPT can even design structures for different incident angles and polarization states, as well as simultaneous design for multiple angles requirements. In this way, OptoGPT solves all previously mentioned challenges in multilayer inverse design, significantly pushing the research boundaries forward.

## Discussion and Outlook

In conclusion, this review summarized the approaches of different types of inverse design algorithms specifically for multilayer thin film structures, starting from traditional optimization-based methods, to the recent popular deep-learning based algorithms. Compared to other types of photonic structures, multilayer structure is special as it can be represented using three different methods. Each of them has its unique advantages and drawbacks, and corresponds to a specific type of inverse design algorithms. It was only until recently that researchers have identified the sequential representation, a unique data type that is specific to layered structure. This new method has shown promising potential to reshape the inverse design.

Despite the great success deep learning-based methods have achieved, there are still remaining questions to be solved. Perhaps the biggest question to deep learning is the “black-box problem”, i.e., the predictions from NN cannot be properly explained due to the high-level complexity and nonlinearity inside NN. In inverse design, this explainability is very important to understand why certain structures are preferred during the design process, e.g., if the designed structure is coming from some existing physical behaviors or potentially includes new physical mechanism that has not been discovered before. However, uncovering the black-box problem has not been extensively explored yet. New ideas and approaches are needed to tackle this issue and reveal hidden physical principles.

Another issue is that many deep-learning based methods requires a large dataset. Although after training, NN are extremely fast for inverse design, generating the training dataset itself can take a long time. This leads to the dilemma of whether the speed-up in inverse design process is worthwhile. In addition to the data-based deep learning, currently another type of neural operator-based learning has been popular in solving complicated partial differential equations, such as PINN[112,113] and DeepONet[114,115]. By integrating the physics rules and mathematical equations into NN, these models can directly learn the complicated mapping from input to output without any data. More efforts are needed in order to incorporate these methods to inverse design.

## Acknowledgments

We thank the National Science Foundation FET-2309403 for the support of this work.

### Author contributions

T. M. and L. G conceived the review framework. All authors contribute to the writing.

### Declaration of interests

The authors declare no competing interests.